# Application of Social Network Analysis in Evaluating Risk and network resilience of Closed-Loop-Supply-Chain

Sara Akbar Ghanadian[a], Saeed Ghanbartehrani[b,*]

**Abstract.**

Closed Loop Supply Chain (CLSC) networks are an attractive topic in both industry and academic research due to their positive environmental effects and waste reduction. Performance evaluation of CLSC networks is challenging due to the variety of facility types and complex relationships among them. In this study, a framework based on Social Network Analysis (SNA) is proposed to evaluate the individual components of a network and identify the critical facilities whose disruptions can affect the entire network. The proposed SNA metrics are applied to a CLSC network case study based on real data and the results and their interpretation are presented. Practical recommendations and actionable guidelines are provided are provided based on the results to mitigate the identified risks and improve the flexibility and resilience of the network. The SNAinSCM R package is developed and shared on GitHub to facilitate the computation and visualization of the discussed metrics.

## 1. Introduction

Supply chain network design problems are perceived as important decision-making problems in supply chain management (SCM) owing to their notable role in the strategic planning process (Aravendan and Panneerselvam, 2016; Pati et al., 2013). Performance evaluation has a key role in the success of organizations through setting objectives, evaluating performance, and composing appropriate action plans. There are various performance metrics for supply chains with respect to order planning, suppliers, service and satisfaction, and cost (Gunasekaran et al., 2004). However, the existing supply chain metrics are not still sufficient to address the complexity of supply chain networks (Abu-Suleiman et al., 2004), and there is still a great deal of confusion in the theory and practice of performance evaluation of supply chains (Bourne et al., 2018). Analyzing different types of facilities and their relationships with other facilities in supply chain networks leads to a deeper understanding of the network nature of supply chains and identifying potential issues disrupting the flows of supplies, information, and services in the network.

Social Network Analysis (SNA) is a powerful method to study the relationship patterns in social networks, telecommunication, and epidemic diseases to name a few (Borgatti, 2005a; Candeloro *et al.*, 2016; Landherr *et al.*, 2010). However, the application of SNA metrics to CLSC networks has remained unexplored in the literature. The objective of this research is to interpret SNA metrics in the context of CLSC networks and to provide a new decision-making tool to evaluate the performance of the individual components of CLSC networks by applying selected node level Social Network Analysis (SNA) metrics. The proposed approach in this research is the first instance of the application of degree centrality, strength centrality, and reducing factor SNA metrics to CLSC network evaluation. The discussed SNA metrics are presented as an SNA based performance evaluation framework that helps the decision-makers to understand the strengths and weaknesses of an existing supply chain network. This framework can also be used to assess the performance of proposed supply chain network designs before the implementation and strategically plan the resources to improve performance and flexibility. An R package titled SNAinSCM package is developed and shared on GitHub (AkbarGhanadian, 2020) to facilitate the calculation and visualization of the SNA metrics on CLSC networks.

## 2. Literature Review

### 2.1. *Closed Loop Supply Chain Network Design*

Closed loop supply chains (CLSC) offer efficient sustainable processes in which the products are recovered using sustainable practices including recycling and remanufacturing in order to decrease environmental degradation while improving the profitability (Neto et al., 2010).

CLSC networks are typically composed of forward and reverse logistics networks. Forward logistics network involves the same activities as traditional supply chains which are delivering products from suppliers to customers including manufacturers, distribution centers, retailers, and customers (Pishvaee et al., 2009). The reverse logistics network is responsible for transferring the returned products for reprocessing operations including recycling, reusing, or remanufacturing (Jayaraman et al., 2003). Reverse logistics is considered environmentally friendly because it offers the possibility of obtaining a product from waste or recycled materials (Dekker et al., 2012). US firms spend up to $100 billion annually on reverse logistics activities involved in the product return procedure (Ghadge et al., 2016). Although CLSCs are more complex compared to the conventional supply chains, they are adopted by many leading companies with large supply chains such as Dell, Walmart, and Apple to improve their recycling efforts.

Network design models for CLSC networks can be classified into separated forward and reverse logistics, and integrated forward-reverse logistics models. While in many studies forward and reverse logistics are considered separately for simplicity, integrated forward-reverse logistics results in better designs with lower cost (Lee et al., 2013). Different types of integrated forward-reverse logistics CLSC networks are modeled in (Pishvaee et al., 2009; Khajavi et al., 2011; De Rosa et al., 2014; Ponce-Cueto and Molenat Muelas, 2015; Ghadge et al., 2016; Yavari and Zaker, 2019) as shown in Table 1.

Table 1: Integrated forward-reverse logistics in CLSC networks literature

| | | Supply chain network components | | | | | | | Models features | | | | | | | | | | | | | | |
|---|---|---|---|---|---|---|---|---|---|---|---|---|---|---|---|---|---|---|---|---|---|---|---|---|
| **V**: Decision variable<br>**E**: Fixed location of centers<br>**S**: Stochastic<br>**D**: Deterministic<br>**P**: Predefined rate<br>**F**: Fuzzy rate<br>**NR**: Not reported in model<br>**#**: Number of facilities | | configuration of the integrated reverse-forward logistic network | | | | | | | Variables | | | | | | Return rate | | | | | Period | | | |
| | | | | | | | | | | | | Enviro. Factor | Demand | | | | | Disposal rate | | # of products | | | | |
| Year | References | Production/ recovery centers | Hybrid facility | Separated collection centers | Disassembly centers | Disposal centers | Retailer/Customer zones | Warehouses | Transportation amount | Facility capacity | Location-Allocation | Carbon emission | Stochastic/Deterministic | Percent of demand | Random | Fuzzy | Interval | Predefined/ Fuzzy | One product | Multi products | One period | Multi period | Facility capacity |
| 2009 | Pishvaee et al. | #5,V | #10,V | | | #3,V | #15,E | | * | | * | | S | * | | | | P | * | | * | | * |
| 2010 | El-Sayed et al. | #3,V | | | #3,V | #3,V | #4,V | #3,V | * | | * | | S | * | | | | | | * | | * | * |
| 2011 | Khajavi et al. | #3,V | #5,V | | | #3,E | #4,E | | * | | * | | D | * | | | | P | * | | * | | * |
| 2011 | Lee et al. | #2,E | #3,V | #3,V | | | #5,E | #3,V | * | | * | | D | * | | | | | | * | | * | * |
| 2013 | Rosa et al. | #6,E | #12,V | #12,V | | | #100,E | #12,V | * | * | * | | D | | | * | | | | * | | * | * |
| 2015 | Ponce-Cueto & Muelas | #NR,E | | | #NR,E | #No,E | #NR,E | #No,V | * | | * | | D | | | | | | | * | | * | |
| 2016 | Ghadge et al. | #3,E | #5,V | | | | #3,E | | * | | * | | D | * | | | | | * | | * | | |
| 2017 | Kang et al. | #4,E | #NR,V | | | #1,E | #30,E | | * | | * | * | D | | * | | | F | * | | * | | * |
| 2019 | Yavari and Zaker | #3,E | | | | | #20 E | #8,V | * | | * | * | D | * | | | | | | * | | * | * |

In integrated forward-reverse logistics networks, hybrid processing facilities offer the same functionality of traditional distribution centers used in forward logistics with the added benefit of collecting the returned products. Utilizing hybrid facilities reduce the cost and environmental footprint owing to sharing the required infrastructure and material handling equipment (Pishvaee et al., 2009). The case study CLSC network in this research benefits from the integrated-forward reverse logistics design.

*2.2. Social Network Analysis*

Social network analysis (SNA) is the process of investigating the patterns of links in a network based on the graph theory (Kim et al., 2011). There has been a growth in the application of social network analysis in many fields of study including trade relationship networks (Mueller et al., 2007), disease epidemic networks (Candeloro et al., 2016), poly-drug trafficking networks (Hughes et al., 2017), and telecommunication networks (Al-Shehri et al., 2017). Although social network analysis (SNA) has been used to study relationship patterns in networks in many disciplines, its application to supply chain networks has remained unexplored.

Several research studies have discussed the application of SNA metrics to supply networks rather than supply chain networks. A supply network is a two side network consisting of supplier and buyer, while supply chain networks include diverse components such as suppliers, production or manufacturing facilities, retailers or customers, and transportation channels to deliver the products to the customers (Kim et al., 2011; Santoso et al., 2005). Borgatti and Li (Borgatti and Li, 2009) performed a theoretical review of centrality metrics including betweenness, degree centrality, and closeness SNA metrics in supply networks. Kim et al. (Kim et al., 2011) applied and interpreted node and network-level degree centrality, closeness, and betweenness SNA metrics in a firm-supplier supply network scenario. Mueller et al. (Mueller et al., 2007) mentioned the potential application of SNA metrics including degree centrality, closeness, and betweenness SNA metrics to supply chains. However, both Kim et al. (2011) and Mueller et al. (2007) failed to provide any measurement and interpretation of the SNA metric values in the context of supply chains or networks. Galaskiewicz (2011) presented their understanding of relevant social network theories to supply chains. Although a general idea of the possibility of application of SNA to supply chain networks was discussed, the objective, application, and interpretation of the metrics were missing.

In this research, node-level SNA metrics including degree centrality (in and out), strength centrality (in and out), and Reducing factors ($R_{disperse}$ and $R_{absorb}$) are measured and interpreted for all types of facilities in forward and reverse logistics in CLSC networks. Using SNA metrics as a decision-making tool in supply chain design and analysis to identify high risk facilities in supply chain networks and more specifically CLSC networks is a novel application of SNA discussed in this study. The collection of the aforementioned SNA metrics forms a new decision-making tool to evaluate the components of CLSC networks to aid in CLSC network design.

*2.3. Degree Centrality*

Centrality is a commonly used metric to evaluate the robustness of social networks (Borgatti, 2005a). Borgatti (2005b) defined node level degree centrality as "the number of links incident upon a node". Easy interpretability of degree centrality metric is the main reason for its extensive application in SNA (Candeloro et al., 2016). In-degree centrality is the number of links a given node receives from other nodes in the network. Out-degree centrality reflects the number of nodes that a given node points to. The total degree can therefore be defined as the number of nodes that a given node both points to and receives from. In social networks, the node with the highest degree centrality is considered a central point of communication and likely a major source of information in the network (Opsahl et al., 2010). Having no connection to other nodes or zero degree centrality indicates an isolated node in terms of communications which is typically considered a weakness (Freeman, 1978). In supply chain networks, the most central node can be strategically used as a major distribution channel for the flow in the network.

*2.4. Strength Centrality*

Strength centrality also referred to as weighted degree centrality is the extension of degree centrality where the weights of links are used in analyzing weighted networks (Barrat et al., 2004). Strength centrality reflects the sum of the link weights connecting a given node to its neighbors (Wei et al., 2011). The interpretation of strength centrality depends on the type of network. In a scientific collaboration network, strength centrality reflects the total number of publications of the scientist associated with each node. The literature review did not find any evidence of the application of strength centrality to supply networks or supply chain networks.

*2.5. Reducing Factor*

Candeloro et al. (2016) proposed a new weighted centrality measure referred to as weighted strength centrality (WSC). Strength centrality reflects the sum of the weights on the links connected to a node, however, it lacks the information related to the number of links or how the weights are distributed among them. This can be an issue in networks where the distribution of the weight (flow) is important. Equal weight distribution has been discussed as a desired feature in other networks such as computer networks (Hu et al., 2012) and power distribution networks in the literature (Baloch, 2013).

To incorporate the effect of the weight distribution, a tuning parameter referred to as the reducing factor is calculated and multiplied by the strength centrality metric. When the weights on all edges connected to a node are equal, reducing factor would be equal to 1.0. Any deviation from the uniform (equal) weight distribution, would decrease reducing factor. We use reducing factor as a standalone measure of the weight distribution in this study.

**3. Research Methodology**

In this research, node-level SNA metrics including degree centrality (in and out), strength centrality (in and out), and Reducing Factor ($R_{disperse}$ and $R_{absorb}$) are applied to all types of facilities in forward and reverse logistics networks. The optimal design from a CLSC network model is used as a case study. The case study model is simple enough to be analytically traceable, but general enough to derive insights and generalize to larger and more complex networks. The node-level analysis is provided for Manufacturer-Distribution Center (M-DC), Distribution Center-Retailer (DC-Re), Retailer-Distribution Center (Re-DC), and Distribution Center-Remanufacturer (DC-RM) relationships in the network. It is worth mentioning that the calculation and interpretation procedure proposed in this research is applicable to any medium and large sized CLSC networks with integrated forward and reverse logistic structure.

### 3.1. Node Level Degree Centrality

Degree centrality is the simplest centrality metric in social network analysis (SNA) and determines the number of direct contacts of a node (Landherr et al., 2010). At the node-level, degree centrality $C_D$ is the count of the neighbors for a node in the network. Assuming an adjacency matrix A with elements $a_{ij} = 1$ when there is an edge from node *i* to *j* and 0 otherwise, in- and out-degree centrality are calculated based on **Equations** **(1)** and **(2)** as follows.

$$\text{Out-degree centrality} \qquad C_{D-out}(x) = \sum_{j=1}^{N} a_{xj} \qquad (1)$$

$$\text{In-degree centrality} \qquad C_{D-in}(x) = \sum_{i=1}^{N} a_{ix} \qquad (2)$$

The degree centrality metric represents the number of relations with other facilities for a given facility in a supply network. A high in- or out-degree centrality reflects high transactional intensity or related risks for the facility the node represents (Kim et al., 2011).

In the CLSC networks, out-degree centrality is applicable to distribution centers and retailers in the forward logistics from which the products are sent out. High in-degree centrality results in the diversity in suppliers which improves flexibility and resilience in case of disruptions and unexpected events (Costantino and Pellegrino, 2010).

In-degree centrality is considered for distribution centers and remanufacturing facilities in reverse logistics. High in-degree centrality in reverse logistics means receiving returned products from more sources. This is challenging for distribution centers since the facility needs to collect, process, store, and ship the returned products. For remanufacturing facilities, high in-degree centrality implies neither negative nor positive interpretation because their objective is to have enough capacity and infrastructure to process returned products rather than having relationships with more distribution centers.

Out-degree centrality is considered for manufacturers and distribution centers in forward logistics. High out-degree centrality means serving more facilities which implies more challenges in ensuring on-time delivery, order processing, responding to changes in demand, and transportation planning. On the other hand, having high out-degree centrality in forward logistics is indicative of being a common supplier with the capability of multi allocations as a manufacturer or distribution center, and lower risk of excess or obsolete inventory because of more relationships with other facilities. However, high out-degree centrality combined with low in-degree centrality in distribution centers is a potential risk in forward logistics. Out-degree centrality in reverse logistics is measured for retailers and distribution centers. Retailers and distribution centers with high out-degree centrality have the benefit of more relations with more processing facilities for the returned products. Similar to the case of forward logistics, high out-degree centrality in reverse logistics is indicative of being a common supplier capable of supporting multiple retailers while such facilities need to be supervised and monitored carefully to ensure they perform optimally in the supply chain. In the case of disruptions in the operation of such facilities, multiple retailers and remanufacturing facilities would be affected.

*3.2. Strength Centrality*

**Equation (3)** shows the strength centrality for a node based on the weight $W_{ij}$ associated with the edge connecting nodes *i* and *j*.

$$S_i = \sum_{j}^{N} W_{ij} \qquad (3)$$

In social networks, the link weights represent the number or frequency of interactions between nodes, geographic distance, or the travel time between them (Wei et al., 2011). In-degree and out-degree strength centrality can be calculated according to **Equations (4)** and **(5)** as follows.

Out-degree strength centrality $$S_{D-out}(x) = \sum_{j=1}^{N} W_{ij} \qquad (4)$$

In-degree strength centrality $$S_{D-in}(x) = \sum_{i=1}^{N} W_{ij} \qquad (5)$$

In CLSC networks, strength centrality is defined for both the origin and destination facilities that interact with one another. The out-degree strength centrality is defined for the origin facilities (i.e. suppliers) and in-degree strength centrality is discussed for the destination facilities.

In-degree strength centrality is considered for distribution centers and retailers in forward logistics. High in-degree strength centrality for distribution centers is typically caused by high demand from retailers and reflects more challenges in managing the flow of incoming products, higher investment in storage and transportation infrastructure, equipment, and staff. Dealing with high volumes of products frames the distribution centers as critical facilities in terms of involvement in the supply chain. Retailers with high in-degree strength centrality (i.e. high demand) are considered critical due to their notable role in the sales and therefore profitability of the supply chain. High in-degree centrality indicates greater demand for that retailer. Such retailers and the distribution centers supporting them form crucial distribution channels in the supply chain and are required to be monitored carefully since disruptions in their operation impact many customers.

In-degree strength centrality is considered for distribution centers and remanufacturers in the reverse logistics. High in-degree strength centrality for distribution centers means receiving a higher volume of returned products which involves more effort to manage the returns and ship them to the remanufacturing or recycling centers and requires a higher investment in storage infrastructure, equipment, and staff. Distribution centers with high in-degree strength centrality are also key components in delivering the majority of the returned products to remanufacturing facilities and reverse logistics. As a remanufacturer, high in-degree strength centrality implies challenges in terms of capacity and infrastructure to process the returned products. Since remanufactures perform activities such as disassembly, cleaning, inspection, repair, and sorting that may not occur in the manufacturing process, they typically require more resources and investment compared to manufacturing facilities with similar in-degree centrality. Also, high in-degree strength centrality for a remanufacturer increases the risks associated with the availability, timing, and quality of the remanufacturing process.

Out-degree strength centrality in forward logistics is applicable to manufacturers and distribution centers. Facilities with high out-degree strength centrality have high in-degree strength centrality too. Therefore, all discussions provided for high in-degree strength centrality are also applicable to high out-degree strength centrality. Facilities with high out-degree strength centrality are considered critical and expected to face more challenges in order management and on-time delivery, monitoring and responding to changes in demand, as well as transportation. Out-degree strength centrality in reverse logistics is measured for retailers and distribution centers. Retailers with high out-degree strength centrality deal with higher volumes of returned products from customers which requires more investment in return logistics and transportation. Distribution centers with high out-degree strength centrality face challenges in handling the logistics of the returned products and transportation to remanufacturing or recycling facilities.

*3.3. Reducing Factor (R)*

Reducing factor is a metric used to compare the weight distribution between nodes having identical degree and strength centrality. The three scenarios presented in Figure 1 have the same degree and strength centrality, while their reducing factors are different.

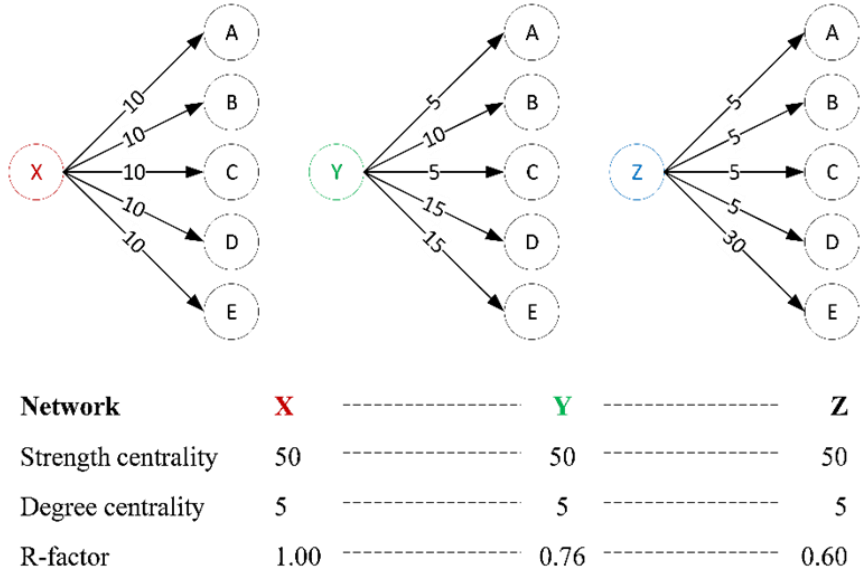

**Figure 1.** The weight distribution in three sample networks with the same number of nodes and total flow

The area under the cumulative function of percentage weight distribution is denoted by $AUC_{Fw}$, and the area under the uniformly distributed percentage weight is denoted by $AUC_{max}$. Reducing factor (R) is the ratio between ($AUC_{Fw}$) and ($AUC_{max}$) as shown in **Equation (6)**.

$$R = \frac{AUC_{Fw}}{AUC_{max}} \qquad (6)$$

It is critical to sort the weights in ascending order before calculating the cumulative weight. The cumulative percentage weight plots for the three networks in Figure 1 are shown in Figure 2.

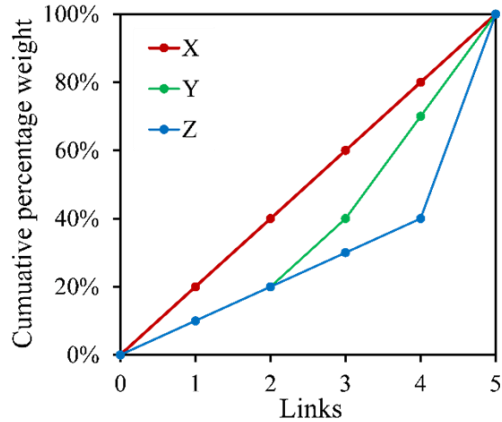

**Figure 2.** Cumulative percentage weight distribution for the three sample networks in Figure 1

We propose two types of reducing factor referred to absorbed and dispersed R factor. For facilities receiving the flow, absorbed R factor ($R_{absorb}$) can be defined as a measure indicating how balanced the flows of the incoming products are. Comparatively, the dispersed R factor ($R_{disperse}$) is defined for facilities sending products to measure the uniformness of the distribution of the products sent. Low reducing factor indicates a significant deviation from uniformly distributed product flows for a node while a close to one value shows a well-balanced flow. The degree centrality needs to be greater than one for a node to be able to calculate $R_{absorb}$ or $R_{disperse}$.

Balanced allocation of customers to distribution centers not only decreases the frequency of backorders and late deliveries but also increases utilization rate and order fill rate in distribution centers (Zhou et al., 2002). Including the balance of the flows in the objective function or constraints in the supply chain network optimization model excessively complicates the solution process while R factor is a simple metric to evaluate and improve the balance of incoming and outgoing flows. Balanced flow of products between facilities result in better utilization of transportation, loading, and shipping infrastructure which reduces the transportation costs and simplifies planning. Unbalanced flows increase the complexity of logistics and planning as well as investment in transportation infrastructure.

## 4. Case Study

A CLSC network with integrated forward-reverse logistics based on real data is used as a case study in this research.

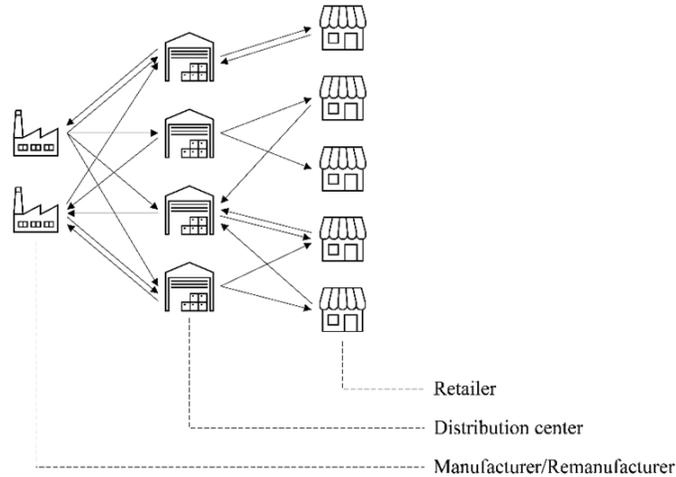

**Figure 3.** CLSC network case study with integrated forward-reverse logistics.

The case study network presented in Figure 3 consists of 5 manufacturing facilities, 50 retailers, and 10 distribution centers. 3 out of 5 manufacturers are capable of both manufacturing and remanufacturing. The distribution centers are hybrid facilities that can send and receive supplies and returns to and from the manufacturing facilities and retailers. Each retailer is assumed to receive supplies from only one distribution center. The distribution centers can be potentially located at any of 118 selected Walmart store locations in the state of Ohio (Holmes, 2011), while the manufacturing facility and retailer locations are considered to be fixed. The demand for each retailer is considered to be constant and known. The return rate from each retailer is 10% of the demand. Manufacturing and remanufacturing capacities for each manufacturer are 1,100,000 and 200,000 respectively.

The shortest network distances between the existing facilities and the candidate locations were calculated using the road network data for the state of Ohio ("OpenStreetMap," 2013). PostgreSQL (PostgreSQL contributors, 2020), PostGIS (PostGIS Project Steering Committee (PSC) et al., 2020), and pgRouting (pgRouting contributors, 2020) were used to perform the required spatial analysis. The mathematical model was solved in CPLEX to find the optimal network design that minimizes the total transportation cost. The solution includes the locations of the distribution centers, as well as the allocation in forward and reverse logistics between the manufacturing/remanufacturing facilities, distribution centers, and retailers. The optimal forward and reverse logistics networks are visualized in Figure 4. The thickness of each edge is proportional to the number of products transferred between the facilities.

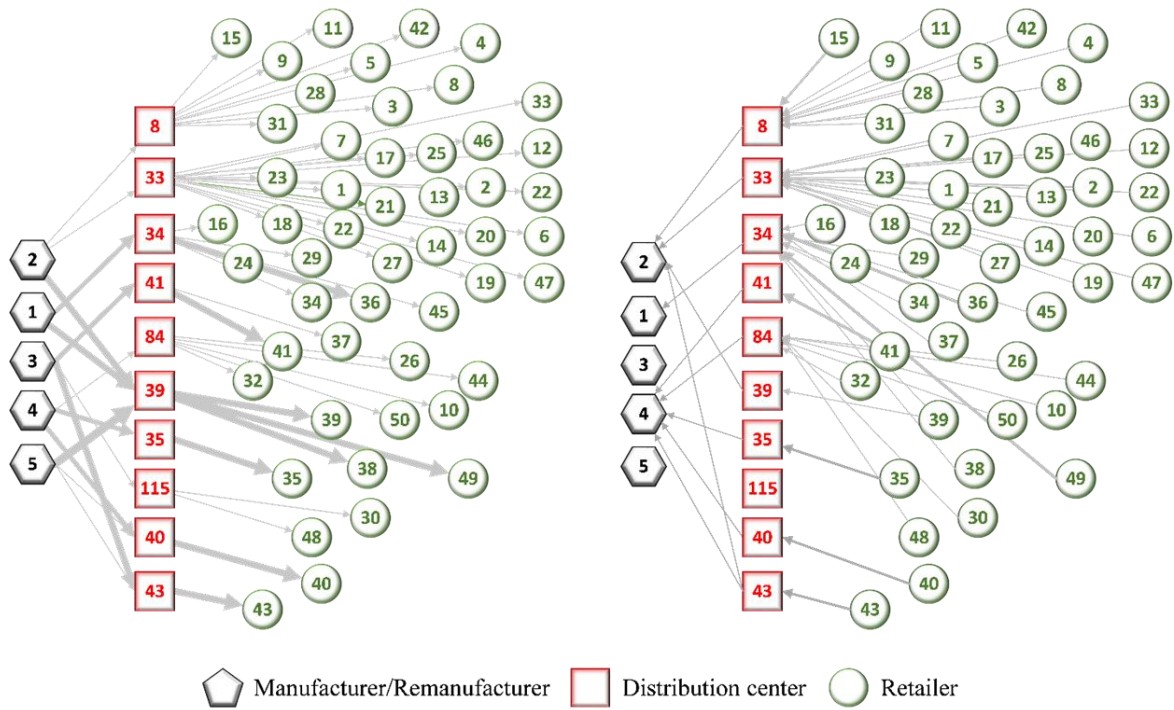

**Figure 4.** The optimal forward (left) and reverse (right) logistics case study networks. The number on each node represents the location of the facility.

In the next step, the discussed SNA metrics are applied to the optimal network design and the results are analyzed.

## 5. Results

SNA metrics including degree centrality, strength centrality, and reducing factor are implemented in SNAinSCM R package to analyze the weighted adjacency matrices for the network design depicted in Figure 4. This R package can be used on any supply chain network and can be installed in Rstudio using the following script.

```
install.packages("SNAinSCM")
devtools::install_github("Saraghanadian/SNAinSCM")
```

The SNA metrics and their interpretation for all facilities will be discussed in forward and reverse logistics networks.

*5.1. Application of SNA Metrics in Forward Logistics Network*

***Manufacturers.*** The out-degree centrality, out-degree strength centrality, and $R_{disperse}$ for manufacturers in forward logistics network are presented in Table 2.

**Table 2.** Forward logistics SNA metrics for manufacturers. (M: Manufacturers, $C_{D\text{-}out}$: Out-degree centrality, $S_{D\text{-}out}$: Out-degree strength centrality, $R_{disperse}$: Dispersed reducing factor)

| M | $C_{D\text{-}out}$ | $S_{D\text{-}out}$ | $R_{disperse}$ |
|---|---|---|---|
| 1 | 2 | 1100000 | 0.90 |
| 2 | 3 | 615794 | 1 |
| 3 | 3 | 1100000 | 0.68 |
| 4 | 3 | 1100000 | 0.68 |
| 5 | 3 | 1100000 | 0.52 |

Results show that manufacturers 2, 3, 4, and 5 have the highest out-degree centrality among the manufacturers.

In supply chain networks, the manufacturer with the highest out-degree centrality is considered the most central manufacturer in terms of involvement and volume of shipped products in the network. Table 2 indicates that manufacturers 1, 3, 4, and 5 have equally high strength centrality utilizing their maximum manufacturing capacity which is 110,000 in the network. It is important to note that manufacturer 2 has some redundant capacity to supply in case of disruptions in the operation of other manufacturers. In supply chain networks, $R_{disperse}$ reflects the balance of flow to other facilities. Manufacturers with $R_{disperse}$ value of 1.0, send equal volumes of products to all distribution centers they are supplying which is a positive point. Table 2 indicates that Manufacturer 2 has $R_{disperse}$ equal to 1.0 while manufacturer 5 with $R_{disperse}$ equal to 0.52 has the worst balance of flow among all manufacturers.

***Distribution Centers.*** The out-degree centrality, out-degree strength centrality, and $R_{disperse}$ for distribution centers in forward logistics network are presented in Table 3.

**Table 3.** Forward logistics SNA metric for distribution centers. (DC: Distribution center, $C_{D\text{-}in}$: In-degree centrality, $S_{D\text{-}in}$: In-degree strength centrality, $R_{absorb}$: Absorbed reducing factor, $C_{D\text{-}out}$: Out-degree centrality, $S_{D\text{-}out}$: Out-degree strength centrality, $R_{disperse}$: Dispersed reducing factor)

| DC | $C_{D\text{-}in}$ | $S_{D\text{-}in}$ | $R_{absorb}$ | $C_{D\text{-}out}$ | $S_{D\text{-}out}$ | $R_{disperse}$ |
|---|---|---|---|---|---|---|
| 8 | 1 | 70176 | NA | 10 | 70176 | 0.78 |
| 33 | 1 | 160684 | NA | 19 | 160684 | 0.75 |
| 34 | 1 | 656348 | NA | 6 | 656348 | 0.28 |
| 35 | 1 | 569499 | NA | 1 | 569499 | NA |
| 39 | 3 | 1708497 | 0.81 | 3 | 1708497 | 1.00 |
| 40 | 2 | 569499 | 0.65 | 1 | 569499 | NA |
| 41 | 1 | 598845 | NA | 2 | 598845 | 0.55 |
| 43 | 2 | 569499 | 0.73 | 1 | 569499 | NA |
| 84 | 1 | 47681 | NA | 5 | 47681 | 0.57 |
| 115 | 1 | 65066 | NA | 2 | 65066 | 1.00 |

In supply chain networks, in-degree centrality for distribution centers shows the number of relationships with manufacturers. Table 3 indicates that distribution center 39 with in-degree centrality of 3 is the most central among all distribution centers. Furthermore, distribution centers 8, 33, 34, 35, 41, and 84 are supplied by only one manufacturer which can be a potential risk in case of disruptions in the operation of the manufacturers.

In supply chain networks, in-degree strength centrality for distribution centers reflects the level of involvement of distribution centers in the network based on the volume of products received from manufacturers. According to Table 3, distribution center 39 has the highest in-degree strength centrality equal to 1,708,497 units. The strength centrality for distribution center 39 constitutes 34% of total products supplied to all distribution centers, while distribution centers 8, 84, and 115 receive a very small portion (less than 2%) of the total products supplied by the manufacturers. This situation makes distribution center 39 a critical distribution center in terms of the level of involvement in the manufacturer-distribution relationship which needs to be taken into consideration when planning resources, staff, and contingencies.

In supply chain networks, $R_{absorb}$ for distribution centers indicates the balance of the received products flows from the manufacturers. In Table 3, $R_{absorb}$ for distribution centers 8, 33, 34, 35, 41, 84, and 115 is not reported since R factor is not applicable to distribution centers with in-degree centrality less than 2. Distribution center 39 has the highest $R_{absorb}$ of 0.81 which is a positive factor. Distribution centers 43 and 40 have lower $R_{absorb}$ values, which is not concerning because they have relatively lower in-degree centrality values. Table 3 indicates that distribution center 33 is the most central distribution center with 19 relationships with retailers. Distribution centers with high out-degree centrality are sensitive facilities because many retailers are affected by disruptions in their operation (i.e. 19 retailers are affected in case distribution center 33 is disrupted).

In supply chain networks, out-degree strength centrality for distribution centers reflects their level of involvement in the network in terms of the volume of products sent to retailers. Bar charts in Table 3 indicate that distribution center 39 supplies the highest volume of products (1,708,497 units) to the retailers which accounts for 34% of the total retailers' demand. Distribution center 39 is a sensitive distribution center from the distribution center-retailer relationship perspective which calls for continuous monitoring and performance evaluation.

Distribution centers 39 and 115 have $R_{disperse}$ values equal to 1.0 having perfectly balanced flows of products supplied to their retailers (see Table 3). $R_{disperse}$ for distribution centers 35, 40, and 43 with out-degree centrality values of 1 is not reported in Table 3 since an out-degree centrality of at least 2 is required to comment on the distribution of the flows. Distribution center 34 has the lowest $R_{disperse}$ value among all distribution centers.

*Retailers.* SNA metrics including in-degree centrality, in-degree strength centrality, and $R_{absorb}$ are reported for distribution center-retailer relationship in forward logistics.

All in-degree centrality values for retailers are one since the case study network is a single allocation model (i.e. each retailer cannot be assigned to more than one distribution center). In multi-allocation models, this measure can have values other than one. Retailers with high in-degree strength centrality (i.e. high demand) are considered critical due to their notable role in the total revenue of the network. Also, high variance in in-degree centrality for retailers is typically a challenge since high-demand facilities are overutilized and sensitive while low-demand facilities are often underutilized. $R_{absorb}$ cannot be reported for any of the retailers since their in-degree centrality values are equal to one, however, this metric can be reported in multi-allocation models.

*5.2. Application of SNA Metrics in Reverse Logistics Network*

*Remanufacturers.* Manufacturing facilities are referred to as remanufacturers when processing the returned products. Three metrics of in-degree centrality, in-degree strength centrality, and $R_{absorb}$ are discussed for remanufacturers.

In reverse logistics network, high in-degree centrality for remanufacturers implies more relationships with distribution centers. Receiving returned products from many distribution centers is typically a challenge. As shown in Table 4, remanufacturer 2 has the highest number of relationships (5) with distribution centers.

**Table 4.** Reversed logistics SNA metrics for manufacturers. (RM: Remanufacturers, $C_{D-in}$: In-degree centrality, $S_{D-in}$: In-degree strength centrality, $R_{absorb}$: Absorbed reducing factor)

| RM | $C_{D\text{-in}}$ | $S_{D\text{-in}}$ | $R_{absorb}$ |
|---|---|---|---|
| 1 | 1 | 200000 | NA |
| 2 | 4 | 101603 | 0.70 |
| 4 | 5 | 200000 | 0.74 |

High in-degree strength centrality typically requires higher investment in infrastructure for processing the returned products. Although recycled or remanufactured material and products can often be used in the manufacturing process, the economy of recycling highly depends on the industry and recycling costs. **Error! Reference source not found.**Table 4 indicates that remanufacturers 2 and 4 have the highest in-degree strength centrality in reverse logistics. These two remanufacturers utilize their maximum remanufacturing capacity which is 200,000 units.Error! Reference source not found.Table 4 indicates that remanufacturers 2 and 4 have $R_{absorb}$ values of 0.70 and 0.74 respectively which indicate reasonably balanced flows of returned products from the distribution centers. A well-balanced flow of returned products is typically less expensive to ship using the same transportation system that ships the products to the distribution centers. A poorly balanced flow may require extra investment on the reverse logistics. No $R_{absorb}$ is reported for remanufacturer 1 since it receives returned products from one distribution center.

*Distribution Centers.* The three SNA metrics of out-degree centrality, out-degree strength centrality, and $R_{absorb}$ are reported for distribution centers involved in Distribution Center-Remanufacturer (DC-RM) relationship while in-degree centrality, in-degree strength centrality, and $R_{disperse}$ are reported for distribution centers in Retailer-Distribution Center (Re-DC) relationship in reverse logistics network.

Table 5 indicates that distribution center 33 has the highest number of relationships with retailers, which is equal to 19. Distribution center 8 and 34 have the second-highest in-degree centrality. Although high in-degree centrality highlights the significance of distribution centers in reverse logistics, such distribution centers have more challenges in on-time delivery, order processing, responding to demand changes, and transportation planning.

**Table 5.** Reverse logistics SNA metric for distribution centers. (DC: Distribution center, $C_{D\text{-in}}$: In-degree centrality, $S_{D\text{-in}}$: In-degree strength centrality, $R_{absorb}$: Absorbed reducing factor, $C_{D\text{-out}}$: Out-degree centrality, $S_{D\text{-out}}$: Out-degree strength centrality, $R_{disperse}$: Dispersed reducing factor)

| DC | $C_{D-in}$ | $S_{D-in}$ | $R_{absorb}$ | $C_{D-out}$ | $S_{D-out}$ | $R_{disperse}$ |
|---|---|---|---|---|---|---|
| 8 | 10 | 7021 | 0.78 | 1 | 7021 | NA |
| 33 | 19 | 16081 | 0.75 | 1 | 16081 | NA |
| 34 | 10 | 200000 | 0.39 | 1 | 200000 | NA |
| 35 | 1 | 56950 | NA | 1 | 56950 | NA |
| 39 | 1 | 39422 | NA | 1 | 39422 | NA |
| 40 | 1 | 56950 | NA | 1 | 56950 | NA |
| 41 | 1 | 56950 | NA | 1 | 56950 | NA |
| 43 | 1 | 56950 | NA | 2 | 56950 | 0.81 |
| 84 | 7 | 11279 | 0.58 | 1 | 11279 | NA |

According to Table 5, distribution center 34 has the highest in-degree strength centrality in reverse logistics network. This means that distribution center 34 receives the highest volume of products from retailers which accounts for 39% of the total products returned to all distribution centers. In this network, distribution centers 8, 33, and 84 receive the lowest volume (less than 2%) of retuned products from retailers. High in-degree strength centrality for distribution center 34 results in more challenges in managing the flow of incoming products. Given the fact that this distribution center also has the highest involvement with respect to out-degree strength centrality in the forward logistics network, decision-makers should consider higher investment in storage and transportation infrastructure, equipment, and staff in this distribution center.

Distribution centers with high out-degree centrality have the benefit of more relations with remanufacturers. Table 5 indicates that distribution center 43 has the highest number of links to remanufacturers which is equal to 2, while other distribution centers are connected to one remanufacturer. Table 5 suggests that distribution center 34 returns the highest volume of products to remanufacturers which accounts for 42% of the total returned products. Processing high volumes of returned products increases the transportation, packaging, sorting, loading, and unloading operations and the costs associated with them. This renders the distribution center 34 critical from the DC-RM relationship perspective and in need of continuous monitoring and performance evaluation.

$R_{absorb}$ for distribution centers 35, 39, 40, 41, and 43 cannot be reported since they have in-degree centrality values equal to one. Distribution center 8 has the most balanced incoming flow with $R_{absorb}$ value equal to 0.78. $R_{disperse}$ is only reported for distribution center 43 (0.81) which has an out-degree value greater than 1.0. The high value of $R_{disperse}$ indicates the balance of outgoing flow from distribution center 43 to remanufactures.

***Retailers.*** Out-degree centrality, out-degree strength centrality, and R$_{disperse}$ are SNA metrics applicable to retailers in the reverse logistics network.

Out-degree centrality for all retailers is 1.0 due to the single allocation assumption in the case study network which means all retailers send the returned products to only one distribution center. However, in multi-allocation models, out-degree centrality for retailers can take values greater than 1.0. In multi-allocation models, high out-degree centrality is indicative of being a common retailer with the capability of shipping returned products to multiple distribution centers.

Out-degree strength centrality reflects the volume of the returned products from each retailer. In cases where returned products indicate defects or quality issues, retailers with high out-degree centrality are areas of concern that need to be planned for. Finally, R$_{disperse}$ is not applicable to nodes with in-degree centrality less than two as discussed earlier. In multi-allocation models, a well-balanced flow of returned products can be easier to manage using the same transportation system used to deliver the products to the distribution centers.

*5.3. Key Facilities Based on the SNA Metrics*

In this section, high-risk facilities in the network based on the discussed SNA metrics are identified and recommendations are provided to mitigate the risks. Risk is defined as the unreliability and uncertainty associated with resources which can cause interruptions and other negative consequences in a supply chain network (Tang and Nurmaya Musa, 2011). Since in the real-world facilities are not always available, it is required to plan for disruptions caused by unavailability in the network design (Jabbarzadeh et al., 2018).

Manufacturers 3, 4, and 5 are critical. Each of these three manufacturers serves three distribution centers which is the highest out-degree centrality in the network while having high strength centrality equal to their maximum manufacturing capacity. High out-degree and strength centrality increases the risk of machine availability and maintenance and results in challenges in on-time product delivery. Furthermore, manufacturer 4 requires extra investment in monitoring and resource management due to its remanufacturing function in addition to its critical role in forward logistics. On the other hand, planning the manufacturing facilities to operate at their maximum capacity for extended periods of time is challenging. A highly structured capacity planning and production scheduling approach is required to make the most efficient use of the available production capacity. A constantly high utilization rate severely limits the production flexibility for the critical manufacturers which is a significant disadvantage. Manufacturing flexibility is the capability of the manufacturing system to deal with changes (Gupta and Goyal, 1989) in machines, processes, products, production volume, expansion, or operations (Parker and Wirth, 1999).

Manufacturer 5 with $R_{disperse}$ equal to 0.52 has the most unbalanced allocation flow among all manufacturers which results in complicated transportation planning.

29 out of 50 retailers (58%) in the network are being supplied by distribution centers 8 and 33. Having the highest out-degree centrality among all distribution centers combined with the lowest in-degree centrality renders these two distribution centers as potential sources of risk in the forward logistics network. Supplying to the majority of the retailers in the network while receiving from only one manufacturer can be seen as a major point of failure in the network. Furthermore, distribution centers 8, 33, and 34 receive 76% of the returned products from most of the retailers in the network and send them to manufacturers 1 and 2 to be remanufactured. Their significant role in reverse logistics, exposes these distribution centers to more challenges in on-time delivery, order processing, responding to changes in demand, and transportation planning. Hence, the distribution centers identified as critical, are major sources of risk due to their excessive challenges in warehouse replenishment, transportation planning, scheduling and capacity planning (Hartmut et al., 2002). Such activities are considered mid-term (i.e. weekly or monthly) or short-term (i.e. daily) planning activities and require more involvement and leadership from the plant managers.

Distribution centers 39 and 34 with the highest in-degree and out-degree strength centrality respectively are key facilities in the forward and reverse logistics networks. Distribution center 39 receives the highest volume of products which accounts for 34% of the total flow of products in forward logistics, and distribution center 34 receives the highest volume of returned products equal to 39% of the total returned products in reverse logistics. The two critical distribution centers significantly impact transit time, lead time to retailers, and the choice of transportation mode in the network and therefore require higher investment in infrastructure as well as transportation management, operations planning and scheduling, and warehouse management.

Distribution centers 41 and 83 with $R_{disperse}$ values of 0.55 and 0.57 respectively, have the most unbalanced outgoing flows among the distribution centers in forward logistics while distribution centers 34 and 84 with $R_{absorb}$ values of 0.55 and 0.57 respectively, receive the most unbalanced flows from the retailers in reverse logistics. These distribution centers face challenges in the utilization of transportation, loading, and shipping infrastructure. Finally, distribution center 34 deals with the highest volume of products in both forward and reverse logistics while receiving the most unbalanced flows from retailers in reverse logistics. The critical distribution centers discussed in this section need further observation and special provisions in supply chain management to operate properly in both forward and reverse logistics.

Retailers 35, 36, 38, 39, 40, 41, 43, and 49 have the highest in-degree strength centrality among all retailers in the network. The retailers with high in-degree strength centrality are critical retailers due to their notable role in the total revenue and profitability of the network. Such facilities are considered major sources of risk in forward logistics because they are receiving from only one supplier which is the result of the single allocation assumption in the case study. Out-degree strength centrality for all these eight retailers is equal to their demand multiplied by the return rate in this case study. In a real-world scenario, out-degree strength centrality reflects the actual volume of the returned products from each retailer. In cases where returned products indicate defects or quality issues, retailers with high out-degree centrality are areas of concern that need to be planned for.

Remanufacturer 4 has the highest in-degree and in-degree strength centrality which makes it a critical remanufacturer in reverse logistics. Performing the remanufacturing activities including recycling and reprocessing combined with high involvement in the forward logistics, highlights manufacturer/ remanufacturer 4 as a critical facility that can negatively affect the manufacturing and remanufacturing cycle time and cost. The increase in cycle time and cost is caused by the increase in maintenance, setup, calibration, packaging, loading and unloading operations, and number of staff involved. Also, such critical facilities suffer from low manufacturing and remanufacturing flexibility which can be a source of risk in the supply chain.

## 6. Conclusions

This study presents the first framework to apply SNA metrics to evaluate CLSC networks and provides empirical results from a case study based on real data. The proposed framework can help with understanding the strengths and weaknesses of existing supply chain networks and comparing supply chain network design alternatives before the implementation and strategically planning the resources to improve flexibility and performance.

The flexibility of a supply chain network is defined as the ability of the network to adapt to new conditions when exposed to internal or external uncertainty (Winkler, 2009). Resilience is the property of a supply chain network that allows it to react to unexpected events and recover to the original state while maintaining continuity of the operations (Ponomarov and Holcomb, 2009). Improving flexibility enhances supply chain resilience (Tang and Tomlin, 2008). Identifying the sources of risk and uncertainty is the first step to improve supply chain resilience. Identifying the pinch points (i.e. facilities whose disruptions significantly affect the network) is a pre-requisite to improving the supply chain resilience (Christopher and Peck, 2004). The interpretation of the results from the proposed SNA metrics enable mangers to identify the critical facilities in the network which are more likely to cause disruption or their failure have a significant effect on the network. A CLSC network with 5 manufacturers, 50 retailers, 10 distribution centers, and 3 remanufactures is used as a case study. The potential locations for the facilities are chosen from 118 Walmart store locations in the state of Ohio (Holmes, 2011) and the real road network distances are calculated using data from OpenStreetMap project ("OpenStreetMap," 2013). The model is solved in CPLEX and the result are stored in excel sheets for further analysis in R. In this study, high risk facilities referred to as critical facilities were identified in the network based on the discussed SNA metrics and recommendations were provided to mitigate the risks associated with them. Addressing the risks associated with the critical facilities based on the provided metrics and guidelines enables stronger planning for redundancy in the network as well as allocating and prioritizing resources efficiently and effectively. We believe that using the proposed SNA metrics in supply chain networks results in higher quality strategic and operational planning of facilities and flows of products to improve flexibility and resilience.

The proposed framework is a decision-making tool that can improve the performance of supply chain networks. Most of the existing methodologies in CLSC network design including (De Rosa et al., 2014; El-Sayed et al., 2010; Kang et al., 2017; Khajavi et al., 2011, 2011; Pishvaee et al., 2009; Ponce-Cueto and Molenat Muelas, 2015) mainly focus on minimizing the total cost. Adopting such design approaches results in several optimal and near-optimal alternative designs which are similar in terms of cost, while their performance in the real world may be significantly different. An analytical decision-making tool to identify the strengths and weaknesses of the alternative network designs is the key to select the best alternative. The proposed framework can help with the evaluation of the components in alternative network designs enabling the decision-makers to eliminate the identified risks or consider a higher investment in resources and infrastructure to mitigate the risks.

The proposed framework can also be used in strategic, operational, and tactical level planning for existing supply chain networks. The "SNAinSCM" R package presented in this study (AkbarGhanadian, 2020) facilitates the application of the discussed SNA metrics in supply chain networks. The R package is available on GitHub and calculates and visualizes degree centrality, strength degree centrality, and R factor for all network components including manufacturers, distribution centers, and retailers. More details and examples of the SNAinSCM R package application are provided in the supplementary section.

## 7. Future Work

The case study in this research is based on a CLSC network with integrated forward-reverse logistics and hybrid distribution centers to send the products to retailers and collect the returned products. Although this is the most cost-effective network architecture according to the literature, separate forward and reverse logistics networks is also a common approach in both academia and industry. The application of the proposed framework to CLSC networks with different architecture may not only require modifications to the metrics and calculation methods, but also needs changes in the interpretation of the results and conclusions. The calculation procedures and interpretation of the results are presented in great detail to enable the researchers and practitioners to perform the necessary changes when needed.

Applying the proposed metrics on facility types not discussed in this study including collection centers, disassembly centers, recycling centers, and so on is a potential area for future work. Furthermore, other SNA metrics like eigenvector centrality can be considered for evaluating the facilities in supply chain networks. Although closeness, betweenness, and eigenvector centrality were considered as potential metrics in the early stages of this study, they were not included in the framework due to the scarcity of empirical results on their application and interpretation in relevant real-world scenarios. Introducing practical performance metrics for supply chain network evaluation improves strategic, tactical and operational planning and control in supply chain management.